\documentclass[useAMS,usenatbib]{mn2e}
\usepackage{amssymb,amsmath,epsfig,times, natbib}
\title[Large scale sloshing]{Large scale gas sloshing out to half the virial radius in the strongest cool core REXCESS galaxy cluster, RXJ2014.8-2430}
\author[S. A. Walker et al.]{S. A. Walker,$^1$\thanks{Email: 
    swalker@ast.cam.ac.uk} A. C. Fabian,$^1$ J. S. Sanders$^2$ \\
  $^1$Institute of Astronomy, Madingley Road, Cambridge CB3 0HA \\
  $^2$Max-Planck-Institute fur extraterrestrische Physik, 85748 Garching, Germany \\
}
\date{}

\begin{document}
\maketitle
\begin{abstract}
We search the cool core galaxy clusters in the REXCESS sample for evidence of large scale gas sloshing, and find clear evidence for sloshing in RXJ2014.8-2430, the strongest cool core cluster in the REXCESS cluster sample. The residuals of the surface brightness distribution from the azimuthal average for RXJ2014 show a prominent swirling excess feature extending out to an abrupt surface brightness discontinuity at 800 kpc from the cluster core (half the virial radius) to the south, which the XMM-Newton observations confirm to be cold, low entropy gas. The gas temperature is significantly higher outside this southern surface brightness discontinuity, indicating that this is a cold front 800 kpc from the cluster core. Chandra observations of the central 200 kpc show two clear younger cold fronts on opposite sides of the cluster. The scenario appears qualitatively consistent with simulations of gas sloshing due to minor mergers which raise cold, low entropy gas from the core to higher radius, resulting in a swirling distribution of opposing cold fronts at increasing radii. However the scale of the observed sloshing is much larger than that which has been simulated at present, and is similar to the large scale sloshing recently observed in the Perseus cluster and Abell 2142.
\end{abstract}


\section{Introduction}
Cold fronts around the central cores of relaxed cool core clusters have been well studied with Chandra (see \citealt{Markevitch2007} for a review), where the high surface brightness allows them to be easily resolved. These cold fronts are believed to be formed due to the sloshing of the cold cluster core as it responds to the gravitational disturbance created by an infalling subcluster's dark matter halo during an off-axis minor merger, as has been simulated by, for example \citet{Tittley2005}, \citet{Ascasibar2006} and \citet{Roediger2011}. The simulations of \citet{Ascasibar2006} predict that the geometric features of older cold fronts should propagate outwards into the lower pressure regions of the cluster as they age. This is because, while the low entropy sloshing gas originally forming the cold front falls back to the centre, it is replaced at the front by higher entropy gas from larger radii. The net result is that the geometric feature of the cold front propagates outwards. However these simulations have only been produced for the central regions of clusters, where there is a steep density gradient and the gas temperature is rising from a cool core.

Large scale gas sloshing in relaxed galaxy clusters has recently been observed for the Perseus cluster (\citealt{Simionescu2012}) and Abell 2142 (\citealt{Rossetti2013}), with cold fronts being observed out to half of the virial radius ($r_{200}$\footnote{$r_{200}$ is the radius within which the mean mass density is 200 times the critical density for a flat universe.}) using ROSAT and XMM-Newton observations, which have a larger field of view than Chandra. The scale of this sloshing is much larger than has been simulated at present, though it is qualitatively similar in terms of its effects on the surface brightness and temperature structure. The large scale sloshing appears to result in a spiral pattern of concentric cold fronts, on opposite sides of the cluster, at ever increasing radius as they age. 

\begin{figure*}
  \begin{center}
    \leavevmode
    \hbox{
      \epsfig{figure=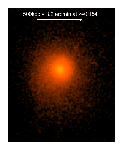,
        width=0.45\linewidth}
         \epsfig{figure=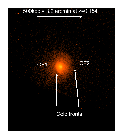,
        width=0.45\linewidth}     
        }
          \hbox{
        \epsfig{figure=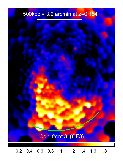,
        width=0.45\linewidth}
         \epsfig{figure=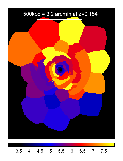,
        width=0.45\linewidth}  
         }       
      \caption{\emph{Top Left}:Exposure corrected, background subtracted, point source subtracted and adaptively smoothed mosaic X-ray image in the 0.7-7.0 keV band from XMM-Newton. \emph{Top Right}: Chandra image of the same region in the 0.7-7.0keV band, showing two central cold fronts which cannot be fully resolved in the XMM image.  \emph{Bottom Left}:Residuals of the 0.7-7.0keV XMM-Newton image after dividing by the azimuthal average, showing the prominent swirling excess to the south. The original image was binned into a Voronoi tesselation with each region containing at least 100 counts. After division by the azimuthal average the image was then smoothed with a Gaussian kernel \emph{Bottom Right}: Temperature map for the same region as the other panels, showing that the southern excess swirl in the surface brightness residuals corresponds to colder gas. The Chandra data have been used for the central 300kpc of the temperature map, and the XMM-Newton data for the regions outside 300kpc. All of the panels have had their coordinates matched.  }
      \label{images}
  \end{center}
\end{figure*}

\begin{figure*}
  \begin{center}
    \leavevmode
    \hbox{
      \epsfig{figure=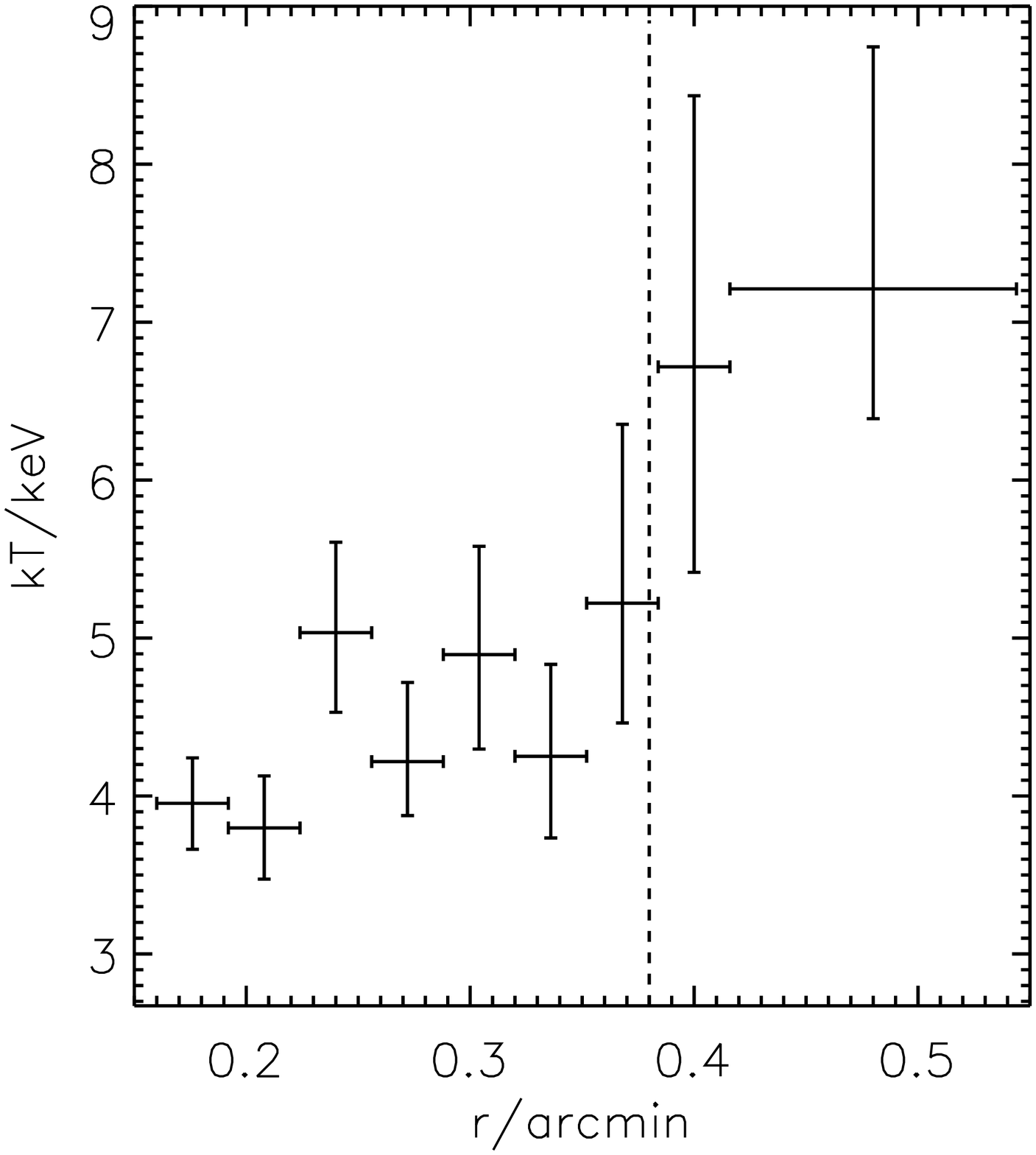,
        width=0.49\linewidth}
         \epsfig{figure=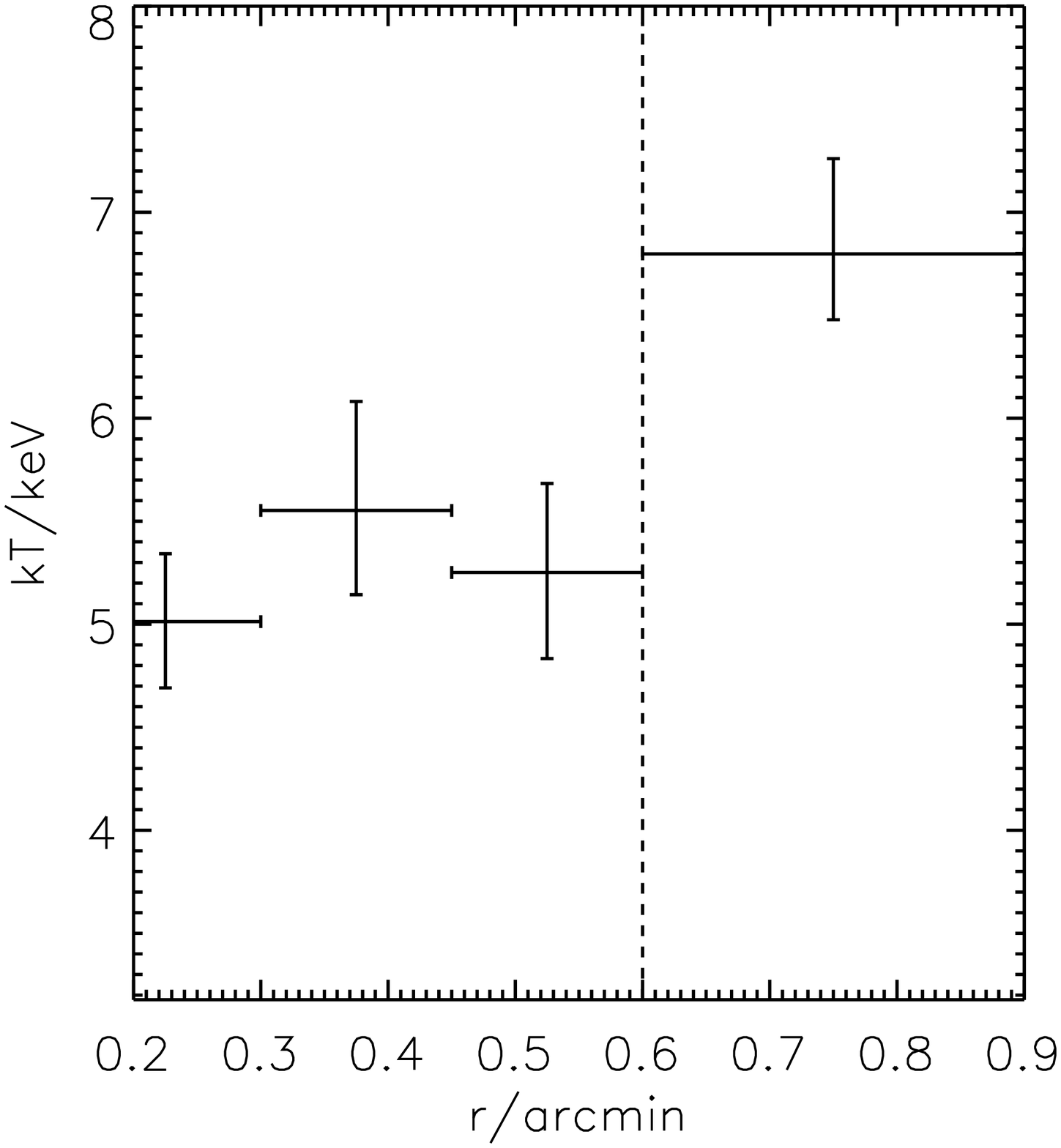,
        width=0.49\linewidth}     
        }
      \caption{\emph{Left}: Temperature  profile across the eastern inner cold front (CF1) observed in the Chandra data.  \emph{Right}: Temperature profile across the western cold front observed with Chandra (CF2). In both cases the positions of the surface brightness discontinuities due to the cold fronts are indicated by the vertical dashed lines.  }
      \label{coldfrontprofiles}
  \end{center}
\end{figure*}

Observations of this gas sloshing at large radii are difficult due to the decrease in the surface brightness of clusters away from the cores, and the relatively small field of view and significant vignetting of the Chandra observations. 
To search for new cases of large scale sloshing in cool core clusters, we therefore searched the cool core clusters of the REXCESS sample of galaxy clusters (\citealt{Boehringer2007}). REXCESS is a representative sample of 31 clusters (with 10 cool cores) at z$\sim$0.1, which spans a wide range in luminosity, temperature and mass. It was designed to prevent bias in X-ray morphology or central surface brightness, and should therefore allow us to begin to understand how common large scale sloshing is for a representative sample. The redshift and masses of the clusters were chosen to ensure that the entire cluster fitted into the XMM-Newton field of view, allowing complete coverage of the clusters to be made, which is ideal for our search for large scale gas sloshing. 

Here we present results from the cool core cluster from the REXCESS sample with the most prominent large scale sloshing features, RXJ2014.8-2430 (hereafter RXJ2014), which is a massive (M$_{500}$=5.4$\times$10$^{14}$ $M_{\odot}$, from \citealt{Pratt2010}) cool core cluster at z=0.154. It is the strongest cool core REXCESS cluster, having the highest central gas density ($n_{e}h(z)^{-2}$ = 0.1291 cm$^{-3}$ at 0.008$r_{500}$) and shortest gas cooling time ($t_{cool}=10^{8.74}$ yr at 0.03$r_{500}$) as found in \citet{Haarsma2010}. 

We use a standard $\Lambda$CDM cosmology with $H_{0}=70$  km s$^{-1}$
Mpc$^{-1}$, $\Omega_{M}=0.3$, $\Omega_{\Lambda}$=0.7. All errors unless
otherwise stated are at the 1 $\sigma$ level.

\section{Identifying RXJ2014 as a large scale sloshing candidate}
\subsection{XMM-Newton data}
We examined the XMM-Newton data for the 10 cool core clusters in the REXCESS cluster sample using the XMM Extended Source Analysis Software (\citealt{Snowden2008}). Images for each cluster were created following the process described in \citet{Walker2013_CentaurusXMM} and described briefly as follows. We ran \textsc{emchain} and \textsc{mos-filter} on the MOS data and \textsc{epchain} and \textsc{pn-filter} on the pn data. Point sources were then removed using the task \textsc{cheese}. Quiescent particle background fields were created using \textsc{mos-back} and \textsc{pn-back}. We then obtained background subtracted, exposure corrected and point source removed images of each cluster in the 0.7-7.0 keV band, mosaicking together the MOS1, MOS2 and pn data using the task \textsc{merge\_comp\_xmm}. 

Next we binned the images into a Voronoi tesselation using the method of \citet{Diehl2006}, with each region containing at least 100 counts, thus reducing the noise in the images. We then divided each image by its azimuthal average from the centre of the clusters, allowing us to observe the residuals to search for the characteristic large scale swirl excess characteristic of large scale sloshing. The cluster RXJ2014 demonstrated the most prominent large scale swirl feature, shown in the bottom left panel of Fig. \ref{images}, extending out to around 800 kpc from the cluster core. The exposure corrected, background subtracted and point source removed XMM-Newton image of RXJ2014 in the 0.7-7.0 keV band is shown in the top left panel of Fig. \ref{images}. We attempted to fit elliptical models to RXJ2014 and examined the residuals from fits to these elliptical models, but the sloshing feature to the south east remained, indicating that it is a robust feature. 

\subsection{Chandra data}

In addition to the XMM-Newton data, we also analysed Chandra ACIS-S data of the central regions of the cluster. The superior, sub-arcsecond spatial resolution of Chandra allows us study the central, small scale features while the larger field of view and effective area of XMM-Newton allows the larger scale sloshing structure to be studied. A detailed understanding of the cluster therefore requires a combination of the two observatories to exploit their complementary strengths. The 0.7-7.0keV Chandra image is shown in the top right hand panel, showing two clear surface brightness discontinuities on opposite sides of the cluster (one at 60 kpc to the east and one at 100kpc to the west), similar to commonly observed cold fronts. 

The details of the observations studied are summarised in Table \ref{obsdetails}.

\section{Spectral analysis}

We binned the Chandra image of the central 300kpc of the cluster into regions containing at least 1500 counts using the contour binning method of \citet{Contbin2006}, which follows surface brightness contours. Spectra were then extracted using \textsc{dmextract}, with response files created using \textsc{mkwarf} and \textsc{mkacisrmf} for each region. The details of the background subtraction and spectral fitting are the same as described in \citet{WangWalker2014}. In short we used stowed backgrounds, scaled to match the 9-12 keV count rate, to remove the particle background. We then used Rosat All Sky Survey (RASS) data for a background annulus around the cluster between 30-60 arcmins to model the components of the background emission, which was modelled as an absorbed powerlaw for the cosmic X-ray background, added to an unabsorbed and absorbed thermal \textsc{apec} component representing emission from the local hot bubble and the galactic halo respectively. The background components were scaled according to the difference in geometric area between the background and foreground regions. 

We then fit each region in \textsc{xspec} in the 0.7-8.0 keV band with an absorbed \textsc{apec} component, allowing the temperature, abundance and normalisation to be free parameters. In all of the spectral fitting we fixed the column density to 13.1$\times$10$^{20}$cm$^{-2}$ as has been found in \citet{Croston2008}, and fixed the redshift to 0.154. All spectral fits were performed in \textsc{xspec} 12.8.1j using the extended C-statistic.

For the XMM data we binned the merged MOS1, MOS2 and pn image into a Voronoi tessellation with each region containing at least 2000 counts. We then followed the spectral extraction and background modelling method of \citet{Walker2013_CentaurusXMM}, which uses the XMM-ESAS software, fitting the MOS and pn data simultanously. Each cluster emission region was then modelled as an absorbed \textsc{apec} component in the same way as the Chandra data. Due to the brevity of the observations it was not possible to spatially map the metal abundance corresponding to the surface brightness features.

\begin{table}
  \begin{center}
  \caption{Observational parameters of the pointings}
  \label{obsdetails}
  
    \leavevmode
    \begin{tabular}{llllll} \hline \hline
    Obs. ID & Observatory & Total clean exposure   \\ 
         &  & per detector (ks)  \\ \hline
    11757 & Chandra & 20  \\ 
      0201902201  & XMM-Newton & 21.5  \\ \hline

    \end{tabular}
  \end{center}
\end{table}

\begin{figure}
  \begin{center}
    \leavevmode
         \epsfig{figure=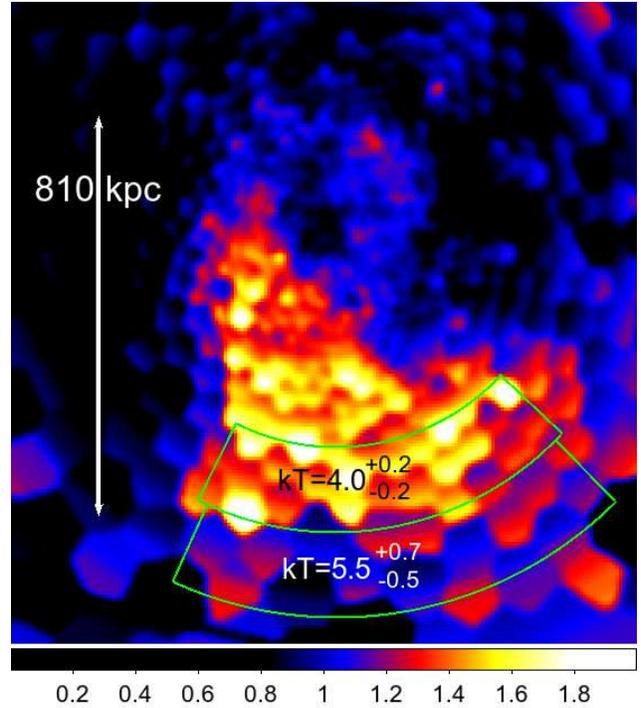,
        width=\linewidth}
    
      \caption{Same as the bottom left panel of Fig. \ref{images}, showing the residuals following division by the azimuthal average surface brightness profile. Spectra were extracted from the XMM-Newton data for the green sectors shown here, inside and outside the proposed cold front. The measured temperatures are shown (in units of keV), with the temperature being higher outside the surface brightness discontinuity in the lower density region, indicating that this is indeed a cold front at large radius and not a shock feature.  }
      \label{CF3_regions}
  \end{center}
\end{figure}

\begin{figure}
  \begin{center}
    \leavevmode
         \epsfig{figure=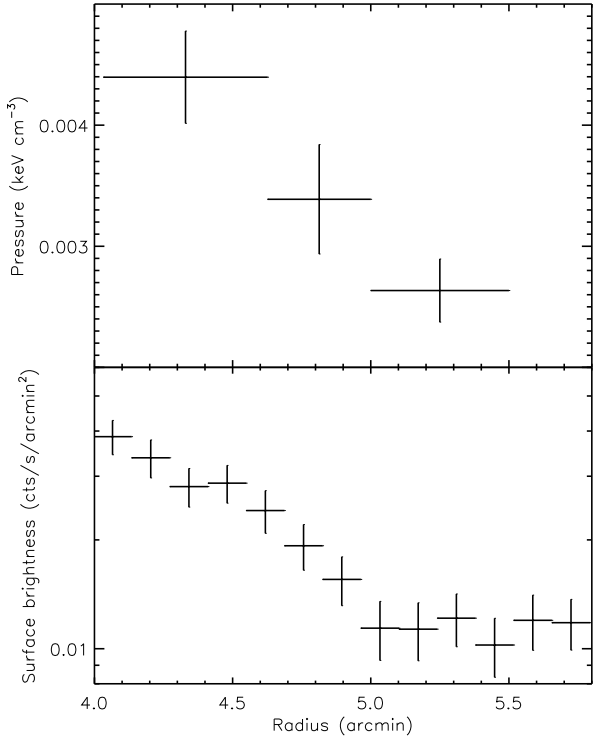,
        width=\linewidth}
    
      \caption{Plots of the pressure (top) and surface brightness (bottom) profiles across the southern cold front, showing the pressure to be continuous across the surface brightness discontinuity.}
      \label{pressure_sb}
  \end{center}
\end{figure}

We combined the resulting Chandra and XMM-Newton temperature maps to produce the temperature map shown in the bottom right panel of Fig. \ref{images}. This allows us to exploit the higher spatial resolution of Chandra in the central regions to map the central cold fronts (CF1 and CF2), while using the XMM-Newton data for the outer regions to exploit its larger field of view and effective area.

\section{Discussion}  

The inner two cold fronts are clearly visible in the Chandra data, indicated as CF1 and CF2 in the top right panel of Fig. \ref{images}. The temperature profiles over these two cold fronts are shown in Fig. \ref{coldfrontprofiles}, showing the abrupt temperature rise outside the surface brightness discontinuities.  

The temperature is also clearly lower to the south of the cluster in the region which coincides with the swirl of excess emission, indicating that we are seeing an old, large cold front (CF3 in the bottom left panel of Fig. \ref{images}) which has risen outwards. This conclusion is strengthened by measuring the ICM temperature immediately inside and outside of the southern excess, using the sector regions shown in Fig. \ref{CF3_regions}. For the sector inside cold front 3, the temperature is 4.0$_{-0.2}^{+0.2}$ keV, while for the sector outside the temperature is 5.5$_{-0.5}^{+0.7}$ keV. This temperature jump outside the surface brightness discontinuity confirms that this is a cold front rather than a shock. Further evidence that this is a cold front is provided by the fact that the pressure profile is continuous across the surface brightness discontinuity, as shown in Fig. \ref{pressure_sb}.

\begin{figure}
  \begin{center}
    \leavevmode
         \epsfig{figure=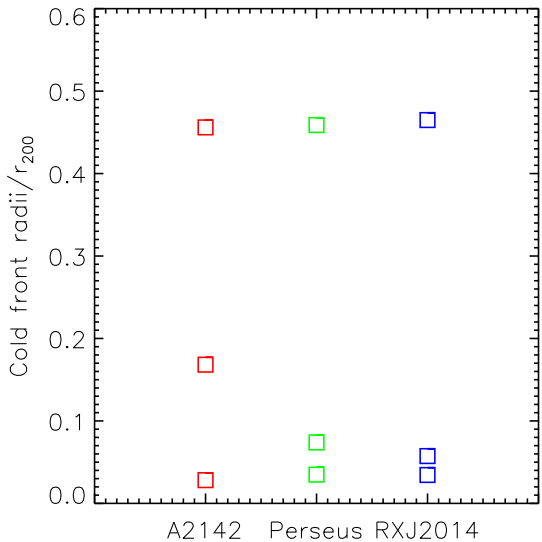,
        width=\linewidth}
    
      \caption{Comparing the cold front locations in RXJ2014 with those in the Perseus cluster and in Abell 2142.}
      \label{cf_loc_compare}
  \end{center}
\end{figure}

The location of the outermost southern cold front is 810 kpc from the cluster core, which is 0.7$r_{500}$ or $\approx$0.5$r_{200}$ (using the value of $r_{500}$=1155 kpc obtained for this cluster in \citet{Croston2008}, and then using the common approximation of $r_{500}$=0.66$r_{200}$). This places the cold front at roughly the same distance from the cluster core as that observed for the outermost cold front in Abell 2142 in \citet{Rossetti2013}, and for the eastern cold front in the Perseus cluster observed in \citet{Simionescu2012}. 

In Fig. \ref{cf_loc_compare}, we compare the radii of the three cold fronts with those observed in the Perseus cluster and in Abell 2142 as a fraction of the scale radius $r_{200}$ to account for the different masses of the clusters. The innermost two cold fronts are closer together than for Abell 2142, and are more similar to those in Perseus. The three cold fronts are at very similar radii to those in the Perseus cluster. Interestingly, the outermost cold front in all three clusters lies at essentially the same scale radius, (0.46$r_{200}$). 

It is most likely that the outer cold front formed from an earlier, different merging event to the one which has formed the inner cold fronts due to the lack of intermediate cold fronts connecting them, and the fact that simulations indicate that the cold fronts should become more difficult to see the further out they are. The similar scaled radii of the cold fronts in RXJ2014, Perseus and Abell 2142 may indicate that the time interval between significant minor merger events in these clusters is similar when scaled appropriately. 

At a redshift of z=0.154, this is the highest redshift galaxy cluster in which large scale gas sloshing has been observed. The fact that RXJ2014 is a strong cool core cluster, with a very low central entropy (1.75$\pm$0.26 keV cm$^{-2}$ , \citealt{Pratt2010}) indicates that the process causing the large scale sloshing is not violent enough to disrupt the cool core. This provides further tests for simulations attempting to reproduce the observed features, which span more than an order of magnitude in distance from the inner cold front at 60kpc, to the outer one at 810kpc.

%\section{Summary}
%
%
%    
%
%\label{summary}

\section*{Acknowledgements}

SAW is supported by the ERC. This
work is based on observations obtained with \emph{XMM-Newton}, an ESA
science mission, and the \emph{Chandra} observatory, a NASA mission.

\bibliographystyle{mn2e}
\bibliography{Sloshing_paper}

%
%\appendix
%\section[]{}
%\label{sec:appendix}
%
%
%
%
%
%
%\clearpage

%\section[]{ROSAT imaging in E11}

\end{document}